\documentstyle[12pt]{article}
\newcommand{\be}{\begin{equation}}
\newcommand{\ee}{\end{equation}}
\newcommand{\ea}{\end{eqnarray}}
\newcommand{\ba}{\begin{eqnarray}}

\begin{document}
%\draft

\thispagestyle{empty}
%\begin{raggedleft}
%IF-UFRJ/xx/97\\
%hep-th/9707204\\
%July/97\\
%\end{raggedleft}
$\phantom{x}$\vskip 0.618cm

%\vfill

%\vfill

\thispagestyle{empty}
%\begin{raggedleft}
%IF-UFRJ/xx/97\\
%hep-th/9707204\\
%July/97\\
%\end{raggedleft}
$\phantom{x}$\vskip 0.618cm

%\vfill

%\vfill

\begin{center}
{\huge Wess-Zumino Terms for the Deformed Skyrme Model}\\
[3ex]{\large C.Neves(1,2) and C.Wotzasek(2)}\\
[3ex]{\em (1)Departamento de F\'\i sica, ICE, Universidade Federal de Juiz de Fora,  
\\36036-330, Juiz de Fora, MG, Brasil,\\
(2)Instituto de F\'\i sica, Universidade Federal do Rio de Janeiro\\
21945-970, Rio de Janeiro, Brazil\\}

%[3ex]{\em Instituto de F\'\i sica\\Universidade Federal do Rio de Janeiro\\
%21945, Rio de Janeiro, Brazil\\}
\end{center}

\begin{abstract}
\noindent A formulation of Skyrme model as an embedded
gauge theory with the constraint deformed away from the spherical geometry is proposed. 
The gauge invariant formulation is obtained firstly generalizing the intrinsic geometry of the model and then converting the constraint to first-class through an iterative Wess-Zumino procedure. The gauge invariant model is quantized via Dirac method for first-class system. A perturbative calculation provides new free parameters related to deformation that improve the energy spectrum obtained in earlier approaches. 
 
\end{abstract}

%\vfill
\newpage

\section{Introduction}

The nonlinear sigma model (NLSM) with a Skyrme term\cite{TS} has received a great deal of attention after the proposal by Adkins et al\cite{ANW} that this is a theory for weakly interacting mesons in the chiral limit resulting from the more fundamental theory for strong interactions, QCD, in the limit when the number of colors $N_c$ is taken very large.
Using the collective coordinate method for the isospin rotation, the authors of \cite{ANW} have performed the semi-classical quantization of the model with the spectrum of static properties being accurate to within $30\%$ of the experimental data, an extraordinary result for such simple model.  This model, however, presents some well known shortcomings.  There is no compelling physical reason to include only the Skyrme term and not higher derivatives potentials. Also the quantum picture is known to be troubled by
operator ordering ambiguities\cite{DW} and the results following
different approaches are not in complete agreement.

Technically the procedure in \cite{ANW} has reduced the field theory problem to that of a free point particle on a $S^3$ sphere.
This theory is an example of a nonlinearly constrained second-class system\cite{PD}. The quantization of such systems has been intensively studied\cite{DW,EG,OS,FH} in various contexts and the relevance
of this problem for the quantization on curvilinear surfaces is well appreciate, both in the path
integral and in the canonical approach.
There has been different proposals in the sense of improving the physical description of the model\cite{RISKY}.
A new possible route, advocated in the present paper, would be to deform away from the intrinsic spherical symmetry.
However there is no conclusive studies of nonlinear constraints of arbitrary geometries available in the literature.
This gap makes it harder to implement this program since the exact shape of the constraint surface able to improve accuracy must come out from a numerical adjustment and is expected to be level dependent.  All this points up to the necessity of a detailed study of quantization over arbitrary constraint surfaces.

It is the purpose of this work to illuminate the quantization of Skyrme's collective rotational mode resulting from quadratic surfaces deviated from the spherical geometry.
While most investigations are done towards understanding the quantum nature directly from the 
2nd-class formulation, we reformulate the model as a gauge theory.     
We provide the gauge invariant reformulation of Skyrme's model for a general nonlinearly
constrained second-class surface.  This is done through the iterative constraint conversion scheme developed by us a few years ago\cite{WN}.

The technique that converses second-class constraints to first-class extending the original phase
space with some auxiliary variables\cite{IJMP}, was suggested by Faddeev and Shatashivilli\cite{F} with the addition
of Wess-Zumino terms (WZ) on the original Hamiltonian.
As far as we know the treatment of nonlinear systems as gauge theories appears already in Balachandran et al in \cite{BALA} and in a number
of papers by Yamawaki et al.\cite{YAMA}.
Our presentation here is inspired by the idea of Kovner and Rosenstein (KR)\cite{KR}, using an analogy with QED to disclose a symmetry hidden in the NLSM.
In \cite{WN} we interpreted the KR symmetry as the gauge symmetry of a Wess-Zumino extended theory, with the geometrical second-class constraints converted to first-class. The hidden symmetry was shown to be a residual symmetry of the WZ orbits. 
The formalism developed in \cite{WN} was used recently in \cite{NW} to study nonlinear models restricted by a more general class of constraints.  It provides a clear-cut geometrical interpretation for the Wess-Zumino gauge symmetry in the extended space as well as in the original configuration space.

In Section 2 we consider, with detailed care, the four dimensional SU(2) nonlinear sigma model with an stabilizing Skyrme term and review the semi-classical expansion of the collective rotational mode.
Reduction to a nonlinear quantum mechanical model depending explicitly on the time-dependent collective variables satisfying a spherical constraint is performed. This will set the stage to allow the Skyrme's model constraint to be deformed and then turned into a gauge theory in the extended WZ phase space.
A systematical treatment of the constraint conversion is developed in the remaining of the paper and is distributed as follows. In section 3 the noninvariant aspect of the theory is reviewed and our notation is introduced. We consider a general setting and study the motion of a point particle in a N-dimensional Euclidean space moving freely on a nonlinear surface $\Omega(q)$ embedded in the $\Re^N$. Special emphasis is given to the {\it symmetric} generating matrix that embraces the geometric information. In section 4 the second-class non-spherical model discussed in section 3 is transformed to a simple gauge invariant theory, written as the sum of the original plus the WZ term. To allow for a concrete calculation we adopt a perturbative point of view and consider small deformations away from the spherical symmetry.  This introduces new parameters into the theory allowing for a better fit with experimental data.
The last Section is reserved to discuss the physical meaning of our findings together with our final comments and conclusions.

\section{The Skyrme Model Revisited}

A few decades ago Skyrme proposed to describe baryons as topological solutions of the NLSM with an appropriate stabilizing term. 
The semi-classical quantization of the model was obtaining in \cite{ANW} separating the collective coordinate.  
Let us consider the SU(2) Skyrmion Lagrangian 

\be
\label{S10}
L'= \int d^3x \left\{ \frac{f_\pi^2}{4} Tr\left(\partial_\mu U^\dagger \partial^\mu U\right) +\frac 1{32 e^2}
Tr\left[U^\dagger \partial^\mu U , U^\dagger \partial^\nu U\right]^2\right\}
\ee
where $f_\pi$ is the pion decay constant and $e$ is a dimensionless parameter. $U$ is a $SU(2)$ matrix transforming as $U\rightarrow AUB^{-1}$ under chiral $SU(2)\times SU(2)$, satisfying the boundary condition $\lim\limits_{r\rightarrow \infty} U= I$ so that the pion field vanishes as $r$ goes to infinity.  There are soliton solutions described by the action (\ref{S10}) whose topological number are identified with the baryon number. To describe the static soliton we start with the ansatz $U(r)=\exp\{i\vec\tau_a . {\hat x}_a f(r)\}$ where $\vec\tau_a$ are Pauli matrices, ${\hat x} = {\vec x}/r$  and $\lim\limits_{r\rightarrow \infty} f(r) =0$ and $f(0)=\pi$. Performing the collective semi-classical expansion in (\ref{S10}), where $U(r,t)= A(t) U(r) A^\dagger(t)$ and $ A \in SU(2)$, we obtain after performing the space integral\cite{ANW},

\be
\label{S20}
L' = -M + {\cal I} \; Tr \left( \partial_0 A \partial_0 A^{-1}\right).
\ee
M and ${\cal I}$ are the soliton mass and the moment of inertia respectively which, in the hedgehog ansatz are given by

\be
\label{S30}
M = 2\pi\!\!\int_0^\infty \!dr  r^2\!\left[  {f_\pi^2}\left(\left({\frac {df}{dr}}\right)^2 + 2 \frac{\sin^2{f}}{r^2}\right) \!+ \frac{\sin^2{f}}{e^2 r^2}\left( 2\left({\frac {df}{dr}}\right)^2 \!+ \frac{\sin^2{f}}{r^2}\right)\right]
\ee
and

\be
\label{S40}
{\cal I} = \frac {8\pi}{3} \int_0^\infty dr \;  r^2 \sin^2{f}\left[ f_\pi^2 + \frac {1}{e^2}\left(\left({\frac {df}{dr}}\right)^2 +  \frac{\sin^2{f}}{r^2}\right)\right] .
\ee
The matrix $A$ may be represented by $A = a_0 + i\; {\vec a}.{\vec \tau}$ satisfying the spherical constraint

\be
\label{S50}
\Omega = \sum_{i=0}^3 a_i^2 - 1.
\ee
In terms of these variables the Skyrmion Lagrangian (\ref{S20}) becomes 

\be
\label{S60}
L' = - M + 2\; {\cal I}\; \sum_{i=0}^3 {\dot a_i}^2 + \lambda \Omega,
\ee
with the spherical constraint (\ref{S50}) being implemented by the Lagrange multiplier $\lambda$.
The Hamiltonian corresponding to (\ref{S60}) is 

\be
\label{65}
H' = M + \frac {1}{8{\cal I}} \sum_{i=0}^3\Pi^2_i - \lambda\Omega,
\ee
with the canonical momenta,

\be
\label{66}
\Pi_i = 4{\cal I} \dot a_i.
\ee
For simplicity we rescale the coordinates and introduce a new Lagrangian and Hamiltonian as

\ba
\label{S70}
L &=& M + L' = \frac 12 {\dot q_i}^2 + \bar\lambda\Omega,\nonumber\\
H &=& -M + H'= \frac 12 p_i^2 - \bar\lambda\Omega,
\ea
where

\ba
\label{S80}
p_i &=& \frac {1}{2\sqrt{{\cal I}}} \Pi_i,\nonumber\\
{\dot q_i} &=& 2\sqrt{{\cal I}}\; {\dot a_i},\nonumber\\
\bar\lambda &=& 4{\cal I}\lambda,\nonumber\\
\Omega &=& q_i^2 - c; \;\;\;\;\; c = 4{\cal I}.
\ea

We have reduced the Skyrmion problem to that of a non-relativistic unity mass particle constrained over a ${\cal S}^3$ sphere, a well known second-class problem. We are now ready to address the question of constraint conversion. A general construction based on the iterative formalism is presented in this paper.  A particular emphasis is given to the extension of the spherical symmetry with dramatic consequences for the energy spectrum.

\section{The Non-Spherical Second-Class Aspect}

In this section we review the main features of a theory for a particle constrained to move on a 2nd-class nonlinear constraint surface.
Let us consider the mechanical O(N) model which is a theory for a point particle with coordinates $q _k\equiv \left( q _1,q _2,q _3, \cdots
,q_N\right) $ moving freely over a nonlinear surface,

\begin{equation}
\label{10}
\Omega = \frac 12 q_k T_{km} q_m - c = 0 .
\end{equation}
Here the surface generator symmetric matrix $T_{km}$ and the yet arbitrary constant $c$ characterize the surface.  We are restricting our study to surfaces bilinear in the coordinates, whose exact structure is encoded in $T_{km}$.  As so these matrices have a decisive role in determining the constraint structure of the model.  This role will become clearer as we progress.  The classical
trajectory and its dynamics is governed by the Lagrangian,

\begin{equation}
\label{20}
{\cal L}=\frac 12 {\dot q}^2-\lambda
\left(\frac 12 qTq-c\right) ,  
\end{equation}

\noindent with $\lambda $ being a Lagrange multiplier enforcing (\ref{10}) as a constraint. To simplify the notation we omit the coordinate indices from now on unless to avoid confusion.  The canonical analysis gives,

\begin{equation}
\label{30}
p_k = \dot q_k ,
\end{equation}
and the primary constraint

\begin{equation}
\label{40}
\phi_1 = p_\lambda \approx 0 .
\end{equation}
The corresponding canonical Hamiltonian reads,

\begin{equation}
\label{50}
{\cal {H}}=\frac 12p_k^2 +\lambda \left(\frac 12 qTq-c\right) ,
\end{equation}
and the primary Hamiltonian is obtained from (\ref{50}) enforcing the constraint (\ref{40}) with a multiplier $u$,

\be
\label{60}
{\cal H}_P = {\cal H} + u \phi_1.
\ee
The nonlinear surface $\Omega(q)$ reappears in the Dirac consistency chain as a secondary constraint,

\ba
\label{70}
\phi_2 &=& \lbrace \phi_1, {\cal H}_P \rbrace\nonumber\\
& = & \Omega .
\ea
The consistency algorithm demands the presence of a tertiary constraint,

\begin{equation}
\label{80}
\phi_3 = pTq
\end{equation}
and a quaternary constraint as,

\begin{equation}
\label{90}
\phi_4 = pTp - \lambda qT^2 q  .
\end{equation}
where $T^2_{km}=T_{kn}T_{nm}$. Constraints $\phi_2$ and $\phi_3 $ have a clear geometrical meaning. The secondary constraint enforces the particle coordinates $q_k$ to take values over the nonlinear surface.  The tertiary constraint, called as transverse, enforces the particle velocity to take values over the tangent space of $\Omega(q)$ at $q_k$.  To disclose the meaning of $\phi_1$ and $\phi_4$ we compute the Poisson algebra that follows from the complete constraint set,

\begin{eqnarray}
\label{100}
\left\{ \phi _1,\phi _4\right\} &=& qT^2q ,\nonumber \\
\left\{ \phi _2,\phi _3\right\} &=& qT^2 q ,\nonumber \\
\left\{ \phi _2,\phi _4\right\} &=&2 qT^2 p,\label{brackets} \\
\left\{ \phi _3,\phi _4\right\} &=&2 pT^2 p + 2\lambda qT^3 q  .\nonumber
\end{eqnarray}
There are no more constraints. From the time evolution consistency for the $\phi_4$ constraint,

\be
\label{110}
0\approx (qT^2 q)\, u +\lbrace \phi_4 , {\cal H}\rbrace  ,
\ee
the multiplier $u$ is determined and the consistency chain stops.  We may use $\phi_4 \approx 0$ to determine the value of the multiplier $\lambda$ and then compute the Dirac brackets,

\begin{eqnarray}
\label{120}
\label{II.40}
\left\{q_k\; ,\; q_m\right\}^* &=& 0\nonumber\\
\left\{q_k\; ,\; p_m\right\}^* &=& g_{km}\nonumber\\
\left\{p_k\; ,\; p_m\right\}^* &=& h_{km}
\end{eqnarray}
where,

\begin{eqnarray}
\label{130}
g_{km}&=&\delta_{km}-t_k\Delta^{-1}t_m\nonumber\\
h_{km} &=& t_k\Delta^{-1}\partial_m t_n p_n -
t_m \Delta^{-1}\partial_k t_n p_n  .
\end{eqnarray}
Here the vector,

\be
t_k=\partial_k \Omega
\ee
and

\be
\Delta=t_k t_k
\ee
are the basic geometric elements of the theory.
The constraints $\phi_1$ and $\phi_4$ have no physical consequence.  Their presence becomes necessary just to eliminate the multiplier sector and
to give consistence to Dirac's algorithm.  In fact, an alternative approach where this sector is eliminate from the outset may be used\cite{BGB,KS}.  This completes the analysis of the 2nd-class aspect of the model.

\section{Quantization of the Non-spherical first-class Skyrme Model}

In this section we present our main result, i.e., the quantization of the deformed gauge invariant Skyrme model and the computation of the energy spectrum.
The conversion of the second-class constraints to first-class by extension of the phase space\cite{IJMP} was originally introduced to avoid the difficulties involving anomalies in chiral gauge theories\cite{F}, by removing the dynamical degree of freedom obstructing gauge symmetry. This scheme has also been used to covariantize the chiral boson constraint\cite{PRL} leading to a system with an infinite chain of 1st-class constraints. This formulation has been of some use in recent developments in the study of superstrings\cite{NB} and dualities\cite{PST}. The logical reasoning behind this approach is to maintain the iterative structure already present in Dirac's formalism. In this sense, the iterative method\cite{IJMP,NW} converts the second-class constraints at their prompt appearance in the consistency chain.

To quantize the deformed Skyrme model and then compute the energy spectrum, it is necessary to know the non-spherical eigenfunctions corresponding to the Hamiltonian operator that now satisfies the new arbitrary geometry. Since it might become so complex, we consider to analyse the Skyrme model perturbatively deviated from the spherical geometry, using the spherical eigenfunctions as a zeroth-order input without any additional incoveniences. To this end we start proposing the perturbative non-spherical Skyrme model as,

\be
\label{40001}
{\cal L} = \frac 12 {\dot q}_i^2 - \lambda( \frac 12 q_i^2 - \frac 12\varepsilon q_i\Delta_{ij}q_j - c),
\ee
where $\varepsilon$ is a small parameter and the surface generator symmetric matrix $T_{ij}$ is assumed to be

\be
\label{40010}
T_{ij} = \delta_{ij} + \varepsilon\Delta_{ij},
\ee
with matrix $\Delta_{ij}$ embracing the deviation from the spherical surface. The Hamiltonian computed through Legendre transformation is,

\be
\label{40020}
{\cal H} = \frac 12 p_i^2 + \lambda( \frac 12 q_i^2 - \frac 12\varepsilon q_i\Delta_{ij}q_j - c).
\ee
As in section 2, the model has four second-class constraints,

\ba
\label{40030}
\phi_1 &=& p_\lambda,\nonumber\\
\phi_2 &=& \frac 12 q_i^2 - \frac 12\varepsilon q_i\Delta_{ij}q_j - c,\nonumber\\
\phi_3 &=& q . p + \varepsilon q . \Delta . p,\\
\phi_4 &=& p^2 - \lambda( \frac 12 q_i^2 + \frac 12\varepsilon q_i\Delta_{ij}q_j) + \varepsilon p . \Delta . p - \lambda(\varepsilon q_i\Delta_{ij}q_j + \varepsilon^2 q_i\Delta_{ij}q_j).\nonumber
\ea
Recall that the first and the last constraints have no geometrical meaning. They are in fact artificial relations generated by the Dirac iterative process to preserve the original consistency presente in the dynamical phase space. Hence, these four second-class constraints could be reduced to only two, the geometrical ones ($\phi_2,\phi_3$).  Computing the Lagrange multiplier $\lambda$ from the last constraint in (\ref{40030}), to first-order in  $\varepsilon$ gives,

\be
\label{40040}
\lambda = \frac {p^2}{2c} \left(1 - \frac {\varepsilon}{2c}q . \Delta . q + \varepsilon\frac{p . \Delta . p}{p^2}\right).
\ee
Bringing back this result into the Hamiltonian (\ref{40020}), it becomes,

\be
\label{40050}
{\cal H} = \frac {1}{4c} q_j^2p_i^2 + \varepsilon\left[ \frac {p^2}{4c} (q . \Delta . q) - \frac {p^2}{8c^2} (q . \Delta . q)q_j^2 + \frac {p^2}{4c} (q . \Delta . q) - \frac 12 p . \Delta . p +       \frac{\varepsilon}{4c}(p . \Delta . p)q_j^2\right],
\ee
with two second-class constraints, denoted subsequently as $\chi_1=\phi_2$ and $\chi_2=\phi_3$. At this stage the gauge invariant formulation of the second-class model with the introduction of the canonical WZ variables $(\theta,\pi_\theta)$ starts\cite{IJMP,NW}. Firstly changing the nature of the geometrical constraint from second to first-class,

\ba
\label{40055}
\chi_1\rightarrow\tilde\chi_1 &=& \chi_1 + \theta\nonumber\\
\chi_2\rightarrow\tilde\chi_2 &=& \chi_2 - \pi_\theta,
\ea
so that the Poisson bracket between them becomes,

\be
\label{40060}
\lbrace \tilde\chi_1,\tilde\chi_2 \rbrace = 0.
\ee

After that, it is imperative to compute the new Hamiltonian that preserves the Dirac iterative process in the extended phase space,

\be
\label{40070}
\lbrace\tilde\chi_1,\tilde {\cal H}\rbrace = \left[\frac{q^2}{2c} + \frac{\varepsilon}{2c}(q . \Delta . q)\right]\tilde\chi_2.
\ee
Therefrom the Hamiltonian counter-term is found and, subsequently, the gauge invariant Hamiltonian,

\ba
\label{40080}
\tilde{\cal H} = \frac{q^2p^2}{4c} &+& \varepsilon\left[ \frac {p^2}{4c} (q . \Delta . q) - \frac {p^2}{8c^2} (q . \Delta . q)q_j^2 + \frac {p^2}{4c} (q . \Delta . q) - \frac 12 p . \Delta . p + \frac{\varepsilon}{4c}(p . \Delta . p)q_j^2\right]\nonumber\\
 - \frac {1}{4c}(q^2 &+& \varepsilon(q .\Delta . q))(q^2 + 2\varepsilon (q .\Delta . q))\pi_\theta^2.
\ea
This concludes the gauge reformulation of the deformed Skyrme model.

To conclude our task, this gauge invariant model has to be adequately quantized. Thereby, the WZ symmetry is partially fixed as,

\be
\label{40090}
\psi = \theta.
\ee
Consequently, $\tilde\chi_2$ turns to second-class constraint,

\be
\label{40100}
\lbrace\theta,\tilde\chi_2\rbrace = q^2 + 2\varepsilon q . \Delta . q
\ee
that allows to compute the corresponding Dirac brackets,

\ba
\label{40110}
\lbrace q_i, q_j\rbrace^* &=& \lbrace p_i, p_j\rbrace^* = 0,\nonumber\\
\lbrace q_i, p_j\rbrace^* &=& \delta_{ij}.
\ea
Notice that these brackets turn out to be canonical, illustrating the Maskawa-Nakajima theorem\cite{MN}. The reduced Hamiltonian is rewritten as,

\be
\label{40120}
\tilde{\cal H} = \frac{q^2}{4c} p_iM_{ij}p_j + \varepsilon\left[ \frac {p^2}{2c} (q . \Delta . q) - \frac {p^2}{8c^2} (q . \Delta . q)q^2 + \frac {q^2}{4c} (p . \Delta . p) - \frac 12 p . \Delta . p - (q . p) \frac{q . \Delta . p}{2c} + \frac{(q . p)^2}{2cq^2} (q . \Delta . q)\right]
\ee
where the matrix $M_{ij}$,

\be
\label{40130}
M_{ij} = \delta_{ij} - \frac {q_iq_j}{q^2}
\ee
is singular. $\tilde\chi_1|_{\theta=0}=\chi_1$ remains a first-class constraint, identified as the ``Gauss Law" generator of the remaining symmetry.

At this point we are interested to compute the energy spectrum of the model from the invariant Hamiltonian above and to compare it with the one given by ANW\cite{ANW}. To calculate the energy spectrum we will use the Dirac quantization method for first-class constraint systems. To this end we impose the first-class constraint in the spatial sector of the Hilbert space,

\be
\label{40140}
(\hat q^2 - c)|phys>=0.
\ee
The momenta representation must reflect the constraint presence and its associated algebra. The first-class nature of the geometric constraint is represented as,

\be
\label{40150}
[\hat q^2,\hat p_i] = 2 \imath\hbar\hat q_i.
\ee
This algebra is satisfied by the following commutator,

\be
\label{40160}
[\hat q_i,\hat p_j] = \imath\hbar\delta_{ij},
\ee
with the canonical representation for the momenta operator $\hat p_i$,

\be
\label{40170}
p_i = -\imath\hbar \partial_i,
\ee
which satisfies the relations (\ref{40150}-\ref{40160}). The spectrum of the theory is obtained by evaluating the Hamiltonian meam value with typical eigenfunctions defined, for instance, $|polyn> =\frac {1}{N(l)} (q_1 + i q_2)^l$\cite{ANW} to give,

\be
\label{40180}
\hat{\tilde{\cal H}} |polyn> = E_l|polyn>.
\ee
A direct calculation produces in this case

\be
\label{40190}
\hat{\tilde{\cal H}} |polyn> = \frac {1}{4c}l(l + 2)\left\{\left[1 - \varepsilon\frac{q . \Delta . q}{c}\right](q_1 + i q_2)^l + 2\varepsilon(q_1 + i q_2)^{l - 1} q_b\Delta_{bc}\partial_c(q_1 + i q_2)\right\}.
\ee
Therefore, the consistence between (\ref{40180}) and (\ref{40190}) requires,

\be
\label{40200}
(q_1 + i q_2)^{l - 1} q_b\Delta_{bc}\partial_c(q_1 + i q_2) = (q_1 + i q_2)^l.
\ee
For instance, a simple choice for the matrix $\Delta_{cb}$, turning the surface slightly oblate,

\be
\label{40210}
\Delta = \pmatrix{0 & 0 & 0 & 0\cr 0 & 1 & 0 & 0\cr 0 & 0 & 1 & 0\cr 0 & 0 & 0 & 0}
\ee
produces the following eigenvalue,

\be
\label{40220}
E_l = \frac {1}{4c} l(l+2)\left[1 + 2\varepsilon\left(1 - \frac{q_1^2 + q_2^2}{2c}\right)\right].
\ee
Clearly, the extra term reflects the correction generated by the deformation away from the spherical symmetry. Note that this deformation matrix $\Delta_{ij}$ has produced a new free parameter that may be used to better the fit energy spectrum to experimental data.
Notice that for $\varepsilon > 0$ the above result improves the ANW estimate

\be
E_l= E^{ANW}_l + \varepsilon \Delta E_l
\ee

Before closing this section we would like to mention an important feature related to the internal consistency of the formalism.  If we choose  the deformation matrix function $\Delta_{ij}=\delta_{ij}$ (which automatically satisfy (\ref{40200})) then such deformation corresponds only to a change in the radius of the sphere.  A direct calculation from (\ref{40190}) shows that the spectrum remains unchanged.  This is important since the radial direction was seen to be the gauge orbit\cite{NW} - spheres of different radii are physically equivalent.  Only deformations on the geometry produce energy corrections.

\section{Final Discussions}

The quantization of the Skyrme model is a well known example of quantum mechanics on curved space.
This comes from the presence of nonlinear second-class constraints but the passage to the quantum world is plagued by ambiguities.
In general the use of the Poisson bracket structure is in conflict with the constraints.
The use of the Dirac brackets helps solve this problem classically, but the new algebraic structure
ends up being coordinate dependent posing a new problem in terms of ordering ambiguities
at the quantum level. Different representations for the momentum operator brings additional terms into the quantum Hamiltonian associated to the zero-point motion. The quantum dynamics becomes dependent on the particular ordering adopted and is not unique.

To cure these sort of problems we propose to treat nonlinear second-class systems
as gauge theories by converting the constraints into 1st-class.
After conversion has been accomplished, the quantization may be implemented by the Dirac or Gupta-Bleuler
procedures where the 1st-class gauge constraints select the physical Hilbert space.
Since the Poisson brackets remain as the underlying algebraic structure, no ordering ambiguities
affect the quantization process.
The question that remains is how to convert efficiently the constraints so that different geometries could also be treated.
This seems important to improve the quantization of the rotational modes of the Skyrme theory.
The development of a formalism to answer this question is the main contribution of this work.

The development of this formalism was done in Section 4. We studied the model with the most general quadratic constraint and showed that a single WZ term produces the desired covariantization of the Skyrme Lagrangian, even for arbitrary geometries.
The exact computation of the spectrum for an arbitrary geometry has shown to be a too difficult problem.  We proposed to handle it in a systematic fashion by perturbative technique.  A variety of deformations may then be examined through a proper choice of the matrix $\Delta_{ij}$ satisfying the condition (\ref{40200}).  The new parameters related to the geometry deformation modify the energy spectrum.  Notice that when these parameters preserve the spherical symmetry the old results of ANW are recovered.  It should be noticed that this deformation occurs in the internal space of constraints and, therefore, can not be compared directly with the results of the ``deformed skyrmions" proposed earlier where the deformation was encoded in the profile function of the hedgehodge solution\cite{RISKY,urbana,LM}
Our solution provides new free parameters related to the deformation which may be adjusted adequately to fit experimental results.

\end{document}